\documentclass[superscriptaddress,twocolumn,prl,showpacs]{revtex4}

\usepackage{amsmath,amssymb}
\usepackage{graphics}
\usepackage{epsfig}

\begin{document}

\title{Suppression of molecular decay in ultracold gases without Fermi statistics}

\author{J. P. D'Incao}
\affiliation{Department of Physics and JILA, University of Colorado,  Boulder, Colorado 80309-0440, USA}
\affiliation{Department of Physics, Kansas State University, Manhattan, Kansas 66506, USA} 
\author{B. D. Esry}
\affiliation{Department of Physics, Kansas State University, Manhattan, Kansas 66506, USA} 

\begin{abstract}
We study inelastic processes for ultracold three-body systems in which only one 
interaction is resonant. We have found that the decay rates for weakly bound molecules
due to collisions with other atoms can be suppressed not only without fermionic 
statistics but also when bosonic statistics applies. In addition, we show that at ultracold temperatures 
three-body recombination involving a single resonant pair of atoms
leads mainly to formation of weakly bound molecules which, in turn, are stable against 
decay. These results indicate that recombination in three-component atomic gases can be used as 
an efficient mechanism for molecular formation, allowing the achievement of high molecular densities.
\end{abstract} 
\pacs{34.50.-s,34.50.Cx,67.85.-d,31.15.x}
\maketitle 

In recent years, the efficiency of diatomic molecule formation in ultracold quantum gases and their stability once 
formed have become key ingredients for many experiments. For example, the remarkable stability of weakly 
bound molecules in ultracold gases of fermions in two different spin 
states \cite{LongLivedMol} has greatly helped experimental studies in the BEC-BCS crossover regime 
\cite{BECBCS-Studies}
when tuning the two-body $s$-wave scattering length $a$ from $a>0$ to $a<0$ near a Feshbach resonance. 
The long lifetimes recently observed for ultracold heteronuclear molecules \cite{RbK} might also
pave the way for future studies on ultracold polar molecules. The stability of weakly bound molecules where at least   
one atom is a fermion has been interpreted as a result of suppression due to the fermionic statistics 
governing both  atom-molecule and molecule-molecule collisions \cite{Petrov,OurPapers}. 
When bosonic statistics plays a role, however, weakly bound molecules tend to decay rapidly, limiting the 
experimental possibilities for studies near a Feshbach resonance \cite{OurPapers,Paul,BraatenVrel}. 
Obviously, besides stability, efficient production of molecules is also of crucial experimental importance.

In this Letter, we present a series of new predictions that opens up new possibilities for obtaining
long molecular lifetimes and that also provide an alternative path for efficient molecular formation.
Surprisingly, we have found that weakly bound molecules of two resonantly interacting atoms can still 
be stable against collisions with other atoms even when bosonic statistics applies.  
Specifically, inelastic collisions between these weakly bound molecules and {\em any} 
atom not resonant with either of the molecule's atoms
is suppressed as $a^{-1}$, therefore allowing long molecular lifetimes near a Feshbach resonance. 
This suppression in the absence of fermionic statistics has recently been observed in Ref.~\cite{RbK} 
in an ultracold mixture of weakly bound $^{87}$Rb$^{40}$K molecules and $^{40}$K atoms in an nonresonant state. 
Here, we show that the mechanism that determines the $a^{-1}$ suppression can be traced to the Efimov physics 
governing the atom-molecule interactions \cite{OurPapers}. 
In contrast to the cases with fermionic suppression \cite{Petrov,OurPapers}, our predicted $a^{-1}$ suppression  
holds for {\em any} mass ratio between the collision partners. 
The sole requirement is that only the atoms bound in the molecule interact
resonantly. 

We will also show that three-body recombination in such systems
displays interesting and potentially useful behavior.
In the one- and two-component bosonic gases and 
in bose-fermi mixtures studied to date, recombination can lead to a substantial fraction of deeply bound molecules 
\cite{OurPapers,OurK3Papers,BraatenReview}, 
giving them in the process enough kinetic energy 
to escape from typical traps. Moreover,  any weakly bound molecule 
formed can eventually decay rapidly due to collisions with other atoms and molecules \cite{K3RudiBosons}. 
The situation is better for fermionic gas mixtures of two spin species: 
molecules are stable \cite{Petrov,OurPapers} and the recombination rate into weakly bound molecules ($\propto Ta^{6}$) 
dominates recombination into deeply bound molecules ($\propto Ta^{2.455}$) \cite{OurPapers}. 
These two ingredients ensured the achievement of highly efficient molecular formation and long molecular 
lifetimes \cite{K3Fermions}. 
However, because recombination for fermionic systems scales as $Ta^6$, efficient molecule formation 
is restricted to relatively high temperatures or large scattering lengths \cite{K3Fermions}. 
In this regime, though, collisions can potentially lead to high molecular dissociation rates 
($\propto T^3a^6$ \cite{OurPapers}), limiting the molecular lifetimes.

As we will show, a three-component gas of atoms with only a single resonant interaction
retains the advantages of these fermionic systems --- efficient formation of weakly bound
molecules and long lifetimes --- with the additional advantage of having a non-zero 
recombination rate at
$T\rightarrow0$, allowing the formation of molecules with much lower temperature which is important to 
explore the quantum degenerate regime. We have found that for these systems 
three-body recombination into weakly bound molecules scales as $a^4$ 
at ultracold temperatures while recombination into deeply bound molecules scales as $a^2$. 
This weaker $a$-dependence ensures that near a Feshbach resonance
recombination will lead mainly to formation of weakly bound molecules. 
As shown in Refs.~\cite{K3RudiBosons,K3Fermions}, these weakly bound molecules can stay trapped 
due to the small energy released, minimizing atomic and molecular losses. 
The most important aspect of molecular formation using this mechanism, however, 
is that even in the absence of fermionic statistics, the molecules formed are stable against decay since 
the decay rate scales as $a^{-1}$. This combination of higher formation rate and lower decay rate 
maximizes molecule production.

We study these ultracold three-body collisions by solving the Schr\"odinger equation
in the adiabatic hyperspherical representation. The three-body inelastic
collision rates are determined by solving the hyperradial Schr{\"o}dinger 
equation given by (in atomic units) 
\begin{equation}
\left[-\frac{1}{2\mu}\frac{d^2}{dR^2}+W_{\nu}\right]F_{\nu}+
 \sum_{\nu'\neq\nu} V_{\nu\nu'} F_{\nu'}=E F_\nu.  
\label{radeq} 
\end{equation}
\noindent
Here, $\mu$ is the three-body reduced mass, $E$ is the total energy, $F_{\nu}$ the hyperradial wave function, and $\nu$ a
collective index that represents all quantum numbers necessary to
label each channel. Since the hyperradius $R$ gives the overall size
of the system, this equation describes the collective radial motion  
under the influence of the effective potential $W_{\nu}$, with
inelastic transitions driven by the nonadiabatic couplings
$V_{\nu\nu'}$ \cite{OurPapers}.  

In the regime where the two-body interactions are resonant, i.e., when 
$|a|\gg r_{0}$ (where $r_{0}$ is the characteristic range of two-body interactions), 
the effective potentials $W_{\nu}$ become universal, 
and several analytical properties of ultracold three-body systems
can be derived. In fact, in our framework \cite{OurPapers} both the $a$ and the $E$
dependence of the three-body collision rates can be determined by simply identifying 
the attractive or repulsive character of $W_\nu(R)$
as determined by the influence 
of Efimov physics \cite{OurPapers}. 
Previous work in ultracold three-body collisions \cite{Petrov,OurPapers,BraatenVrel,BraatenReview} 
has analyzed the influence of Efimov 
physics for systems where at least {\em two} of 
the pairwise two-body interactions are resonant.
In this paper, we will show that Efimov physics still has an important impact 
even when just {\em one} interaction is resonant.

For systems with single resonant interaction, the effective potentials are repulsive in the range $r_{0}\ll R \ll|a|$ 
and can be conveniently parametrized by the coefficients $p_0$ and $p_{\nu}$ \cite{OurPapers}:     
\begin{eqnarray}
W_{0}(R)=\frac{p^{2}_{0}-\frac{1}{4}}{2\mu R^2}
\hspace{0.25cm}\mbox{and}\hspace{0.25cm}
W_{\nu}(R)=\frac{p^{2}_{\nu}-\frac{1}{4}}{2\mu R^2}.
\label{EfPot} 
\end{eqnarray}
For $a>0$, $p_{0}$ is associated with the atom-dimer channel and therefore 
plays an important role in atom-molecule processes. The effective potentials associated 
with $p_{\nu}$ describe three-body continuum channels.
The coefficients $p_{0}$ and $p_{\nu}$ in Eq.~(\ref{EfPot}) can be
determined by assuming a zero-range model potential for the
interatomic interactions (we have confirmed that finite-range two-body interactions give the same results)
and writing the three-body wave function in terms of the Faddeev components \cite{OurPapers}.
Doing this, we have determined that $p_{0}$ and $p_{\nu}$ are 
\begin{equation} 
p_{0}=J+1, \qquad
p_{\nu}=J+3, J+5, ... \label{p0pnu}
\end{equation}
where $J$ is the total orbital angular momentum.
Interestingly, in contrast to systems where two or three interactions are resonant, the strengths 
of the potentials $p_{0}$ and $p_{\nu}$ do not depend on 
either the mass ratio or the permutational symmetry of the resonant pair.

Our derivation of these rates' scaling behavior rests on the observation
that the rate limiting step is tunneling through the potential barriers represented by
Eq.~(\ref{EfPot})~\cite{OurPapers}.  This, together with $p_{0}$ and $p_{\nu}$ from Eq.~(\ref{p0pnu}),
gives the scaling laws shown in Table~\ref{TabRates} for systems with one resonant pair of atoms.
We show the results for $XYZ$ systems  
(three distinguishable atoms) and for $BBX$ systems (two identical bosons and a third distinguishable atom). 
For $XYZ$ systems, we have assumed 
that the $X$-$Y$ pair interacts resonantly and the $Z$ atom is in an nonresonant state. For $BBX$, 
we assumed that only the intraspecies interaction is resonant and, therefore, the $X$ atom is in an nonresonant 
state. Note that the scaling laws for both $XYZ$ and $BBX$ systems are exactly 
the same. 
In the table, we show the first three partial wave contributions to vibrational relaxation
of weakly bound molecules, $XY^*+Z\rightarrow XY+Z$ ($V_{\rm rel}$); to three-body recombination into weakly bound molecules, 
$X+Y+Z\rightarrow XY^*+Z$ ($K^{\rm w}_{3}$); to recombination into deeply bound molecules, 
$X+Y+Z\rightarrow XY+Z$ ($K_{3}^{\rm d}$); and to 
dissociation of weakly bound molecules, $XY^*+Z\rightarrow X+Y+Z$ ($D_{3}$). 
\begin{table}[htbp]
\caption{Energy and scattering length dependence for three-body collision
         rates for systems writh a single resonant interaction. Boldface
         indicates the leading contribution at threshold
         ($k\rightarrow0$ where $k=\sqrt{2\mu E}$ for $K_3$ and $D_3$
         and $k=\sqrt{2\mu(E-E_v)}$ for $V_{\rm rel}$, $-E_v$ is the 
         molecular binding energy). The results are valid for all three-body
         systems with three distinguishable atoms ($XYZ$) or with two identical bosons ($BBX$) with 
         $a_{BB}$ resonant.
  \label{TabRates}} 
\begin{ruledtabular}
\begin{tabular}{cccccccc}
& & \multicolumn{2}{c}{$V_{\rm rel}$} & & $K^{\rm w}_{3}$($D_{3}$),$K^{\rm d}_{3}$ &
\multicolumn{1}{c}{$K^{\rm d}_{3}$ } \\ 
&\cline{2-3}\cline{5-7}  
& $J^{\pi}$ & $E$  & $a>0$ & $E$ & $a>0$ & $a<0$  \\ \hline
 & $0^+$ & {\bf const}
           & \boldmath{$a^{-1}$} 
           & {\bf const}\boldmath{$(k^{4})$}  
           & \boldmath{$a^{4}$},\boldmath{$a^{2}$} 
           & \boldmath{$|a|^{2}$} \\
           & $1^-$ & $k^2$ 
           & $a^{-1}$
           & $k^2$$(k^{6})$ 
           & $a^{6}$, $a^{2}$
           & $|a|^{2}$\\
           & $2^+$ & $k^{4}$ 
           & $a^{-1}$ 
           & $k^{4}$$(k^{8})$ 
           & $a^{8}$,$a^{2}$
           & $|a|^{2}$ \\
\end{tabular}
\end{ruledtabular}
\end{table}

\begin{figure}[htbp]
\includegraphics[width=3.in,angle=0,clip=true]{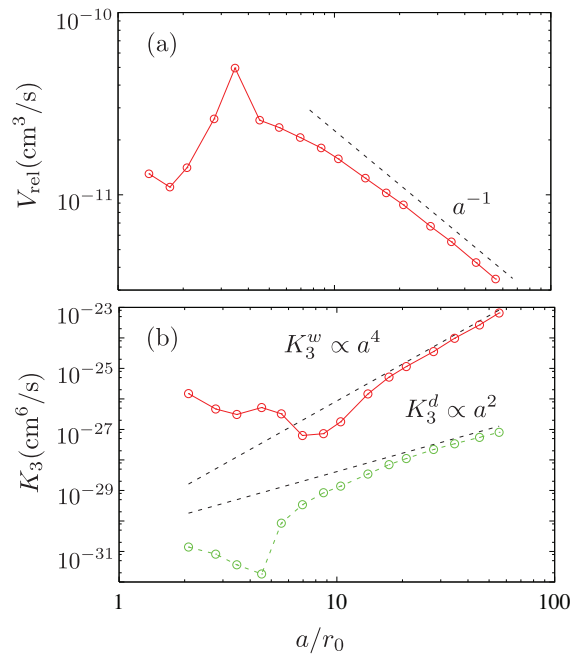} 
\caption{(a) 
Vibrational relaxation rate from weakly bound $XY^*$ molecules
and (b) three-body recombination rates into weakly, $K_3^{\rm w}$,  
and deeply, $K_3^{\rm d}$, bound molecules.}
\label{3bRates}
\end{figure}

Figure~\ref{3bRates}(a) shows our numerical results for $V_{\rm rel}$,
obtained from a direct solution of Eq.~(\ref{radeq}) (see Ref.~\cite{EsryHelium} for details) 
for three distinguishable atoms, $XYZ$, for $J^{\pi}=0^+$ 
($\pi$ is the total parity). We have assumed a finite-range model
two-body interaction in which the resonant 
$X$-$Y$ interaction is adjusted to support one weakly bound 
$s$-wave state whose position controls the scattering length, one deeply bound $s$-wave state,
and one  deeply bound $p$-wave state. Our model $X$-$Z$ interaction 
supports only one $s$-wave state, and, for simplicity, we assumed the $Y$-$Z$ interaction 
has no bound state. The figure shows that when $a/r_{0}\gtrsim10$, our predicted $a^{-1}$ scaling law is verified by the numerical results.
Thus, we have clear evidence that vibrational relaxation is indeed suppressed for large $a$ even in the absence of
fermionic statistics. By extension, this result also support our arguments that
molecular decay in $BB+X$ collisions are also suppressed as $a^{-1}$ 
even though bosonic statistics apply. In a two-component 
gas of $B$ and $X$ atoms, however, long lifetimes for $BB$ molecules can only be achieved if there are no free 
$B$ atoms and with the additional condition of low molecular density, to prevent decay due to 
molecule-molecule collisions.

Recombination for systems involving nonresonant atoms also presents some peculiar
properties. As mentioned previously, for $a>0$, recombination into weakly bound molecules ($K^{\rm w}_{3}$) scales as 
$a^4$ while recombination into deeply bound molecules ($K^{\rm d}_{3}$) scales as $a^2$ 
(see Table~\ref{TabRates}). Therefore, recombination for large $a$ leads mostly to the formation of
weakly bound molecules. This differential scaling with $a$ is in stark contrast to previously studied systems with two or three resonant 
interactions for which recombination is non-zero at $T\rightarrow0$, as 
is the case for one- and two-component bosonic systems and boson-fermion mixtures. In 
these systems, recombination into {\em all} two-body states scales as $a^4$.
In those systems, recombination into the weakly bound state is 
typically dominant as well, but only via the coefficient of $a^4$.
Systems with only a single resonant interaction share this property in addition to the differential scaling with $a$.
Figure \ref{3bRates}(b) shows our numerical calculations of recombination rates for the same system as in 
Fig.~\ref{3bRates}(a). The numerical results support all of the above predictions, including not only the smaller 
magnetude of $K_{3}^{d}$ relative to $K_{3}^{w}$ but also their scaling with $a$. 
Another consequence of this differential scaling is that for $a<0$, i.e., in the absence of weakly bound molecules, 
recombination is 
proportional to $|a|^2$. So, recombination at $a<0$ for systems with a single resonant pair could be suppressed
compared to $a>0$ for large $a$.
\begin{table*}[htbp]
\caption{
  All three-body collision rates relevant for three-component atomic gases. 
  For each mixture, the resonant pair of atoms are indicated 
  by parentheses. The symbol ``$\diamond$''
  indicates  collisions  involving a single resonant interaction.
  \label{TabRates3C}} 
\begin{ruledtabular}
\begin{tabular}{cccccccccc}
Gas Mixture 
&
& 
$V_{\rm rel}^{ij+k}$   
& 
$K^{\rm w}_{3}$, $K^{\rm d}_{3}$ 
& 
&
$V_{\rm rel}^{ij+k}$   
& 
$K^{\rm w}_{3}$, $K^{\rm d}_{3}$ 
&
&
$V_{\rm rel}^{ij+k}$   
& 
$K^{\rm w}_{3}$, $K^{\rm d}_{3}$  \\ \hline
$B_{1}$-$B_{2}$-$B_{3}$
        & $(B_{1}B_{1})B_{1}$ & $a^{}$ & $a^{4}$, $a^{4}$ 
        &  $(B_{1}B_{2})B_{1}$ & $a^{}$ & $a^{4}$, $a^{4}$ &
                                        & &   \\
        &  $(B_{1}B_{1})B_{2}^{\diamond}$ & $a^{-1}$ & $a^{4}$, $a^{2}$  
        &  $(B_{1}B_{2})B_{2}$ & $a^{}$ & $a^{4}$, $a^{4}$ &
                                        & &   \\
       & $(B_{1}B_{1})B_{3}^{\diamond}$ & $a^{-1}$ & $a^{4}$, $a^{2}$ 
        &  $(B_{1}B_{2})B_{3}^{\diamond}$ & $a^{-1}$ & $a^{4}$, $a^{2}$ & 
                                        & &   \\  [0.05in]
$B_{1}$-$B_{2}$-$F$ 
        & $(B_{1}B_{1})B_{1}$ & $a^{}$ & $a^{4}$, $a^{4}$  
         & $(B_{1}B_{2})B_{1}$ & $a^{}$ & $a^{4}$, $a^{4}$ 
        &  $(B_{1}F)B_{1}$     & $a^{}$ & $a^{4}$, $a^{4}$   \\
        & $(B_{1}B_{1})B_{2}^{\diamond}$ & $a^{-1}$ & $a^{4}$, $a^{2}$  
         & $(B_{1}B_{2})B_{2}$ & $a^{}$ & $a^{4}$, $a^{4}$ 
        &  $(B_{1}F)B_{2}^{\diamond}$     & $a^{-1}$ & $a^{4}$, $a^{2}$   \\
        & $(B_{1}B_{1})F^{\diamond}$     & $a^{-1}$ & $a^{4}$, $a^{2}$  
         & $(B_{1}B_{2})F^{\diamond}$     & $a^{-1}$ & $a^{4}$, $a^{2}$ 
         & ${B_{1}F}F$         & $a^{1-2p_{0}}$ & $k^{2}a^{6}$, $k^{2}a^{6-2p_{0}}$   \\  [0.05in]
$B$-$F_{1}$-$F_{2}$
        & $(BB)B$             & $a^{}$ & $a^{4}$, $a^{4}$  
         & $(BF_{1})B$         & $a^{}$ & $a^{4}$, $a^{4}$ 
        &  $(F_{1}F_{2})B^{\diamond}$     & $a^{-1}$ & $a^{4}$, $a^{2}$   \\
        & $(BB)F_{1}^{\diamond}$         & $a^{-1}$ & $a^{4}$, $a^{2}$ 
         &$(BF_{1})F_{1}$     & $a^{1-2p_{0}}$ & $k^{2}a^{6}$, $k^{2}a^{6-2p_{0}}$ 
         & $(F_{1}F_{2})F_{1}$ & $a^{1-2p_{0}}$ & $k^{2}a^{6}$, $k^{2}a^{6-2p_{0}}$   \\
        & $(BB)F_{2}^{\diamond}$         & $a^{-1}$ & $a^{4}$, $a^{2}$  
         & $(BF_{1})F_{2}^{\diamond}$     & $a^{-1}$ & $a^{4}$, $a^{2}$ 
        &  $(F_{1}F_{2})F_{2}$ & $a^{1-2p_{0}}$ & $k^{2}a^{6}$, $k^{2}a^{6-2p_{0}}$   \\  [0.05in]
$F_{1}$-$F_{2}$-$F_{3}$ 
        & $(F_{1}F_{2})F_{1}$ & $a^{1-2p_{0}}$ & $k^{2}a^{6}$, $k^{2}a^{6-2p_{0}}$ & 
                                        & & &
                                        & &   \\
       & $(F_{1}F_{2})F_{2}$ & $a^{1-2p_{0}}$ & $k^{2}a^{6}$, $k^{2}a^{6-2p_{0}}$ & 
                                        & & &
                                        & &   \\
        & $(F_{1}F_{2})F_{3}^{\diamond}$ & $a^{-1}$ & $a^{4}$, $a^{2}$ & 
                                        & & &
                                        & & 
\end{tabular}
\end{ruledtabular}
\end{table*}

The results obtained here for three-body processes involving a single resonant pair --- summarized in 
Table~\ref{TabRates} --- combined with previous results \cite{OurPapers} exhaust all possible three-body processes 
relevant to three-component atomic gases. The competition between the various collision 
processes in a multi-component gas is dictated fundamentally by their energy and scattering length dependence, 
although it is also crucial to take into account the atomic and molecular densities. In fact, 
manipulating the atomic and molecular densities can make favorable a particularly desired 
collisional behavior. 

For completeness, we show in Table~\ref{TabRates3C} 
the three-body collision rates for {\em all} possible ultracold three-component atomic gases. 
For each mixture, we have listed the three-body processes according to which interatomic interaction is resonant by 
indicating the resonant pair of atoms, and we have only included the dominant
contribution at threshold. Three-body processes involving a single resonant pair 
are indicated by ``$\diamond$'' while all the others were obtained in previous work 
\cite{Petrov,OurPapers,BraatenVrel,BraatenReview}.
Futher, we have $2\leq p_{0} \leq 4$ for $V_{\rm rel}$ and 
$0\leq p_{0} \leq 2$ for $K^{\rm d}_{3}$, depending on the mass ratio between the collision
partners \cite{OurPapers}, and we have omitted modulation factors that produce the interference minima 
and resonant peaks related to Efimov physics \cite{OurPapers}. 

We can see from the table that in all possible three-component mixtures weakly bound molecules can be expected to 
be long-lived with relaxation scaling as $a^{-1}$, provided that all free atoms are nonresonant.
In fact, for gas mixtures where distinguishable fermions are resonant, 
weakly bound molecules are expected to be long-lived
irrespective of the identity of the free atoms.

Given the suppression possible in atom-molecule collisions, it might be that molecule-molecule collisions are 
the dominant relaxation mechanism in a three-component gas.
For $BF$ molecules, that are themselves fermions, molecule-molecule 
collisions are suppressed at ultracold temperatures due to their $p$-wave character. For $B_{i}B_{j}$ bosonic 
molecules, however, long lifetimes can only be reached for low molecular densities, in order to 
prevent molecule-molecule collisions. In contrast, for $F_{i}F_{j}$ bosonic molecules, 
relaxation due to molecule-molecule collisions scales as $a^{-s}$ with $s>1$ \cite{Petrov} for mass ratios 
$m_{F_{i}}/m_{F_{j}}>0.11603$ (a condition satisfied for commonly used alkali atoms)  and the main 
molecular collisional decay is due to atom-molecule collisions involving a single resonant 
interaction, scaling as $a^{-1}$.

From Table~\ref{TabRates3C} we see that the 
optimal three-component systems utilize molecular formation
close to an $F_{1}$-$F_{2}$ Feshbach resonance. 
In this case, recombination involving two identical fermions is suppressed at ultracold temperatures  while
recombination in both $B$-$F_{1}$-$F_{2}$ and $F_{1}$-$F_{2}$-$F_{3}$ 
mixtures leads mainly to the formation of weakly bound $F_{1}F_{2}$ molecules. 
The most important property of this system is that after the weakly bound molecules are formed,
molecular loss due to collisions with the remaining atoms is suppressed {\em at least} as $a^{-1}$.
Similar benefits can also be realized near a $B$-$F$ resonance under certain circumstances.
For instance, in $B$-$F_{1}$-$F_{2}$ mixtures, efficient $B$-$F_{1}$ molecular formation could 
be achieved if the number of nonresonant $F_{2}$ atoms is higher than
the $B$ and $F_{1}$ numbers. If in the end all bosons are bound in $BF$ molecules,
then collisions with the remaining fermions is suppressed, ensuring
the desired stability of weakly bound molecules.

To summarize, we have shown that vibrational relaxation of weakly bound molecules can be
suppressed near a Feshbach resonance even in the absence of Fermi statistics.  Remarkably,
the root cause of this suppression can be traced to the same kind of universality that yields
the Efimov effect.  The systems that show this behavior can also exhibit differential
scaling with $a$ of the recombination rate into weakly bound molecules that is finite at zero temperature.
These two results combine to make ultracold three-component gases with a single resonant interaction
an attractive alternative for efficiently producing stable, ultracold molecules.

This work was supported by the National Science Foundation, 
the Air Force Office of Scientific Research, and by the W. M. Keck Foundation.

\end{document}